# Continuous Radio Amplification by Stimulated Emission using Parahydrogen Induced Polarization (PHIP-RASER) at 14 Tesla

Andrey N. Pravdivtsev,*[a] Frank D. Sönnichsen[b] and Jan-Bernd Hövener*[a]

[a] Dr. A. P. Pravdivtsev (0000-0002-8763-617X), Prof. Dr. rer. nat. Prof. J.-B. Hövener (0000-0001-7255-7252)
Section Biomedical Imaging, Molecular Imaging North Competence Center (MOIN CC), Department of Radiology and Neuroradiology,
University Medical Center Kiel, Kiel University (http://www.moincc.de/)
Am Botanischen Garten 14, 24118, Kiel, Germany
E-mails: andrey.pravdivtsev@rad.uni-kiel.de; jan.hoevener@rad.uni-kiel.de

[b] Prof. Dr. rer. nat. F. Sönnichsen (0000-0002-4539-3755)
Otto Diels Institute for Organic Chemistry, Kiel University, Otto Hahn Platz 5, 24098, Kiel, Germany

**Nuclear Magnetic Resonance (NMR) is an intriguing quantum-mechanical effect that is used for daily life medical diagnostics and chemical analysis alike. Numerous advancements have contributed to the success of the technique, including hyperpolarized contrast agents that enables real-time imaging of metabolism in vivo. In physics, hyperpolarization has enabled an NMR RASER using a custom low-field setup and high-Q coils only recently. Expanding on this discovery, we report the finding of a RASER emitting $^1$H NMR signal continuously for more than 10 min at a high frequency of 600 MHz. Full chemical shift resolution is maintained and a linewidth of 2 ppb was achieved. A new simulation of a RASER effect in a coupled two spin-½ system was implemented and reproduced experimental findings. The effect was found using standard equipment only; no dedicated setup is necessary, making the NMR RASER accessible to a wide group of researchers.**

The quest for a continuously emitting, high-frequency, liquid-state radio amplification by stimulated emission (RASER) at MHz frequencies and above is ongoing for more than a decade. Pioneering work demonstrated RASERs at low fields with resonance frequencies from 10 Hz to 50 kHz, based on $^3$He or $^{129}$Xe gases polarized with Spin Exchange Optical Pumping (SEOP),[1,2] for hours[3,4] or days. These techniques were used, among other things, for probing fundamental symmetries.[5]

A 400 MHz RASER was enabled by Dissolution Dynamic Nuclear Polarization (dDNP)[6,7], although only for a single shot experiment: RASER signal bursts were observed for 3 seconds after a hyperpolarized sample was poured into a cavity.[8] Continuous emission of NMR signal is not feasible with this approach.

Closest to the high frequency continuous emission was likely a $^{129}$Xe RASER, where 11, 139 MHz bursts were observed at 11.7 T for 8.5 min. Because the polarization was not refreshed, the amplitude was continuously decaying. Interestingly, the magnetization of the sample was so strong that its Larmor frequency was elevated by 3 Hz at the beginning of the experiment due to strong distant dipolar fields.[9]

In a recent breakthrough, Süfke et al.[10] demonstrated a continuous NMR RASER based on an ingenious combination of innovative hardware[11] and Signal Amplification By Reversible exchange[12] (SABRE): SABRE-RASER.[13] Using a sophisticated setup, a spectral resolution of 0.6 Hz was achieved at ≈ 3.8 mT (≈ 4 ppm); unprecedented resolution of J-couplings was reported. However, no chemical shift resolution was feasible at such low fields.[10,13] Unfortunately, the spin physics involved in SABRE forbid a simple translation of this effect to high frequencies. The reason for this is that the SABRE polarization transfer requires a level anti-crossing (LAC); for $^1$H, this occurs at a field of few mT,[14] where field-dependent Zeeman states and field-independent J-coupling states are matched.[15]

Here, we present the observation of an organic, liquid-state NMR RASER that continuously emits radio frequency signal for more than 10 min. Following a single excitation, the effect was observed at room temperature and 600 MHz (Fig 1A, B).

Similar to the SABRE-RASER[10], this new effect is based on the spin order of parahydrogen ($p$H$_2$) (eq. 1),

$$\hat{\rho}_{pH_2} = \frac{\hat{1}}{4} - \hat{\mathbf{I}}_1 \cdot \hat{\mathbf{I}}_2, \qquad (1)$$

which is continuously supplied to a liquid sample (Fig. 1A).[16]

Elegantly, $pH_2$ has spin 0, hence no magnetization, and is impervious to any excitation in its molecular state. Thus, the reservoir of spin order is not affected by pulses or the RASER effect, which takes place in the same physical location. The production of $pH_2$ is well established, requiring no more than $H_2$ gas flow through a catalyst chamber cooled by cryogens or electronically. Stored appropriately, $pH_2$ is stable for days to weeks. $H_2$ can be purchased or produced on site by electrolysis.

In contrast to SABRE-RASER[10], here, $pH_2$ is permanently incorporated into target molecules via homogeneous hydrogenation (see Scheme 1 in methods). The catalytic activity was chosen so that only a small fraction of all precursor molecules is hydrogenated at a time. This way, the reaction was upheld for an extended period of time. In this instance, hyperpolarized signal was emitted continuously after a single "trigger" excitation pulse of 45°(Fig 1A) and observed for 12 min. The duration of the emission is limited only by $pH_2$ supply, the hydrogenation catalyst and the reservoir of receiver molecules. In current implementation, only $pH_2$ was continuously renewed. Refreshing all constituents is easily accomplished using e.g. a continuous flow setup.

Importantly, we expect that this $pH_2$-induced polarization RASER (PHIP-RASER) is functional at all (high) magnetic fields or in parahydrogen and synthesis allow dramatically enhanced nuclear alignment (PASADENA)[17] conditions. This is the case if the J-coupling between the added hydrogens are weak; for J ≈ 10 Hz and chemical shift difference of 1 ppm, this condition is met at fields > 0.2 T, i.e. at $^1$H frequencies above 10 MHz.

PHIP-RASER effect allows to obtain spectra from much longer acquisition time than usual. Two narrow lines were observed in the positions of the added hydrogens after the Fourier transformation of 11 minutes of the acquired RASER signal (Fig 1B). The line shape was Lorentzian-like with a full width at half maximum of ≈ 2 ppb (≈ 1 Hz, Fig 1C).

Interestingly, the resonances of the thermally polarized solvent acquired directly after the excitation (Fig S2, SI) exhibited much wider lines ≈ 100 Hz, broadened by the strong magnetic field inhomogeneity induced by the $pH_2$ supply (Fig S2). Because of this, it remains unclear, how much of the sample actually emits RASER signal. In the absence of these inhomogeneities and magnetic field drift, the RASER line width would be restricted by spectral resolution and of the order of 0.8 mHz for 11 min. A better homogeneity may be achieved in the future by using more sophisticated techniques to dissolve $pH_2$.[18,19]

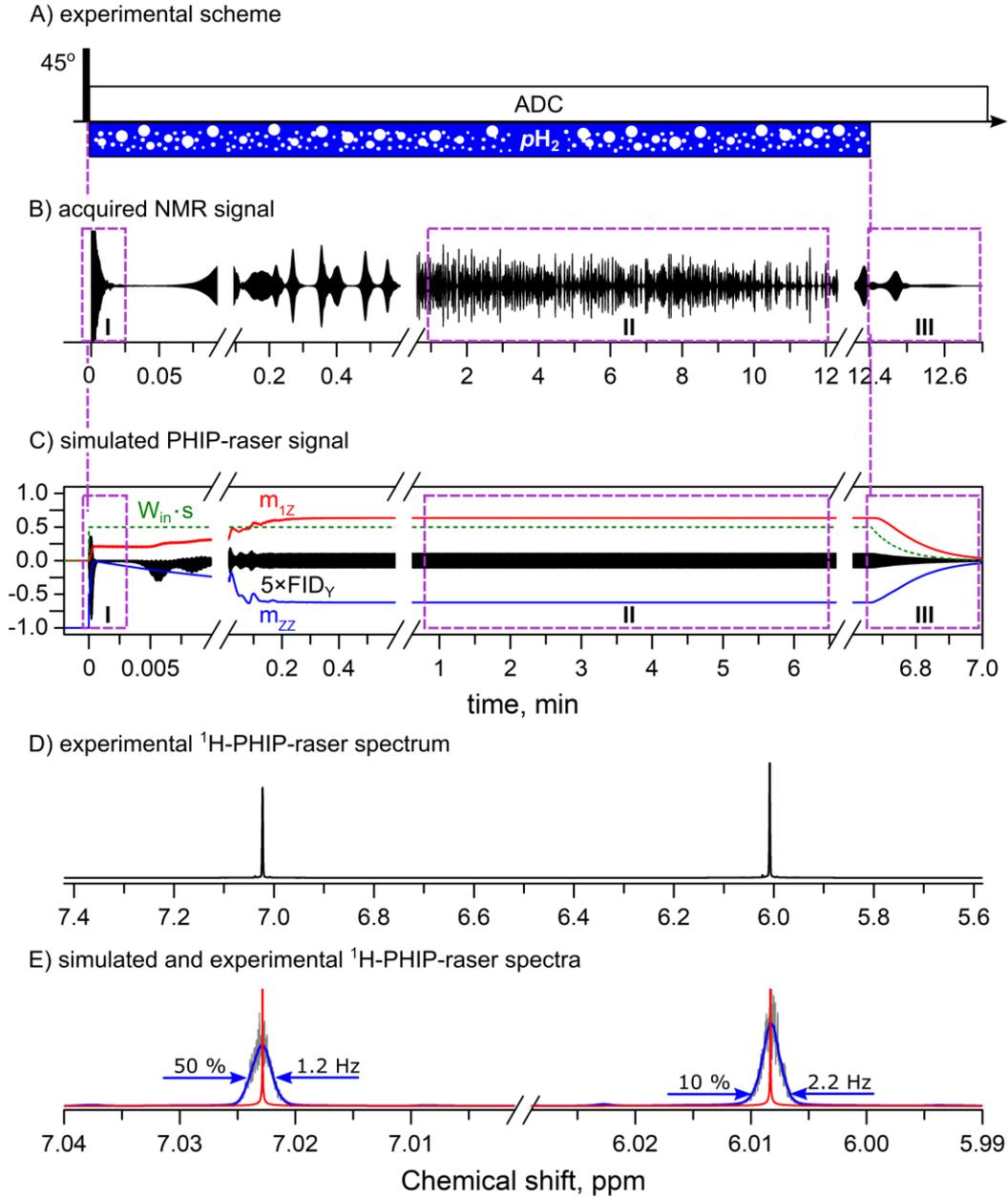

**Figure 1: 12 min of continuous emission of $^1$H NMR signal at 600 MHz.** A liquid state NMR-RASER was induced by supplying $p$H$_2$ into an NMR tube coupled to a resonator in situ (A). Upon one 45° RF-excitation, NMR signal was observed for more than 12 min until the $p$H$_2$ supply was stopped (B). The data was well reproduced by a newly developed quantum mechanical model (C). We found that PASADENA polarization (m$_{zz}$) is converted into longitudinal (m$_z$) and transverse (m$_y$) magnetization by the coupling to the resonator. The Fourier transform of 11 min data (B, dashed box II) exhibited two narrow lines with a full width at half maximum of 2 ppb (D, E; black lines: experimental spectrum, blue: smoothed experimental spectrum, red: simulated spectrum with 10 mHz line broadening).

The interaction between the highly polarized system and RF-cavity is essential for the continuous emission.[3–5,10,11] This interaction reveals itself as radiation damping. Radiation damping rate is defined as $(\tau_{RD})^{-1} = \frac{\mu_0}{4} \hbar \gamma^2 \eta Q c_s |P|$, where $\mu_0$ is a vacuum permeability, $\hbar$ is Planck constant, $\gamma$ is a gyromagnetic ratio, Q is a quality factor of a coil with a filling factor $\eta$, $c_s$ is the concentration of the nuclear spin and $P$ is the longitudinal polarization.[20] This definition of radiation damping rate is very convenient for experiments with large magnetizations in thermal equilibrium that can be described simply with modified Bloch equation.

To elucidate this remarkable effect further, we set out to simulate a coupled two spin-½ system using the density matrix approach and Liouville von Neumann equation.

First of all, it was necessary to introduce a new element into conventional liquid state Hamiltonian of multi-spin system[20] to account for radiation damping:

$$\hat{V}^{rd}(t,\hat{\rho}) = \alpha_{RD} \sum_{k=1,2}(m_{kY}(t)\,\hat{I}_{kX} - m_{kX}(t)\hat{I}_{kY}) \quad (2)$$

This interaction was written in analogy to modified Bloch-Maxwell equation[21] (see more details in SI). Here, $m_{kX,Y}=Tr\left(\hat{\rho}\cdot\frac{1}{2}\hat{I}^{\dagger}_{kX,Y}\right)/Tr\left(\frac{1}{2}\hat{I}_{kX,Y}\cdot\frac{1}{2}\hat{I}^{\dagger}_{kX,Y}\right)$ are amplitudes of transverse polarization of spin $k$ = 1 or 2 of the density matrix $\hat{\rho}$, the amplitudes in general are time dependent. A radiation damping rate without polarization factor, $\alpha_{RD} = \frac{\mu_0}{4}\hbar\gamma^2\eta Q c_s$ was used, which is constant with constant concentration and stable electronic circuitry. On the other hand, $(\tau_{RD})^{-1}$ depends on the polarization, P, which changes continuously because of relaxation or re-hyperpolarization.

We assumed the initial condition of a simplified, weakly coupled, two spin-½ system after hydrogenation with $p$H$_2$ (eq 1) as [16]

$$\hat{\rho}_{PASADENA} = \frac{\hat{1}}{4} - \hat{I}_{1Z}\hat{I}_{2Z} \quad (3)$$

The supply of para-order to the system was implemented with a source operator:

$$\hat{S} = -W_{in}(t)(\hat{\mathbf{I}}_1\cdot\hat{\mathbf{I}}_2) \quad (4)$$

No unity operator was used (see eq 1) to keep the trace over the density matrix equal to 1. $W_{in}(t)$ is a time dependent rate of polarization inflow. Simulations were performed using the MOIN spin library[22,23]; the source code is available online. More detailed description of the theory and simulation is given in SI.

With these additions (eq 2-4), we were able to reproduce the essential effects of the 600 MHz PHIP-RASER (Fig 1C-E). In experiments and simulations alike, a fast decaying signal was observed after the excitation (Fig 1-I); this is the well-known radiation damping effect. After that, the RASER signal occurs in less than few seconds. Here, we find that the experimental signal exhibits much stronger amplitude variations than the simulated one. We attribute this effect to the magnetic field inhomogeneity induced by the $p$H$_2$ supply, which were not taken into account in the simulations.

The same simulations were used to shed light into the classic PASADENA experiment, too. For example, an asymmetric line broadening and emission of "spontaneous", echo-like RASER bursts was observed up to 30 s after the $p$H$_2$ supply was stopped (see Fig S6,8, SI). We found that radiation damping results in conversion of PASADENA spin order (eq 3) into longitudinal, $m_z$, and transverse, $m_{xy}$, magnetization (Fig 1C-I). Experimentally, this is manifested in an asymmetric broadening of spectral lines (Fig S8, SI) and RASER bursts, which may confound the interpretation of the spectra, or causes inefficient polarization transfer to heteronuclei.

These findings establish the PHIP-RASER as the first continuously emitting $^1$H NMR RASER at high field (and high frequencies). Operating with standard NMR equipment, at room temperature and in the liquid state, this PHIP-RASER is likely the simplest implementation of an NMR RASER presented to date. Only minor additions to standard NMR equipment is needed, and the PHIP-RASER is expected to work with most of the installed commercial NMR (and possibly MRI) devices. The key element is a strong coupling between the resonator and hyperpolarized sample.

Moreover, with a frequency range of 10 MHz and above, PHIP-RASER is very flexible. The maximum frequency is currently limited only by the static magnetic fields available (≈ 1.2 GHz).

The long emission > 10 min allows to partially negate the detrimental effects of the strong inhomogeneity present in the sample, as narrow lines of ≈ 2 ppb were observed.

Given a more homogeneous delivery of $p$H$_2$, it appears feasible to further improve on these linewidths. Simulations suggest that extremely narrow lines of a few mHz can be obtained at 600 MHz, effectively breaking the T$_2$ limit. Note that at the same time, the full, 14-T-chemical shift dispersion is maintained. These and other applications yet to discover make PHIP-RASER a highly interesting effect for NMR, physics, chemistry and more.

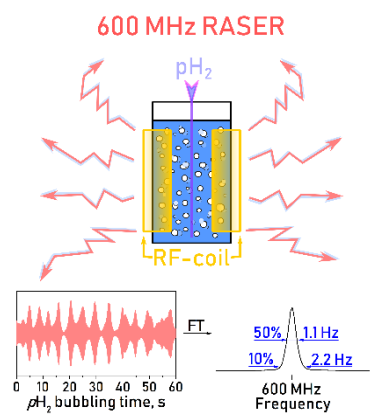

**Acknowledgements**

We acknowledge support by the Emmy Noether Program "metabolic and molecular MR" (HO 4604/2-2), the research training circle "materials for brain" (GRK 2154/1-2019), DFG - RFBR grant (HO 4604/3-1, № 19-53-12013), the Cluster of Excellence "precision medicine in inflammation" (PMI 1267). Kiel University and the Medical Faculty are acknowledged for supporting the Molecular Imaging North Competence Center (MOIN CC) as a core facility for imaging in vivo. MOIN CC was founded by a grant from the European Regional Development Fund (ERDF) and the Zukunftsprogramm Wirtschaft of Schleswig-Holstein (Project no. 122-09-053).

**Keywords:** NMR spectroscopy • Hydrogenation • para-hydrogen induced polarization • PASADENA • RASER

**Supporting information**: Detailed description of the radiation damping theoretical model, and additional experimental and theoretical observations (.PDF) and Bruker acquisition files.

**Author Contributions**

A.N.P. and J.-B.H. designed experiment and wrote the manuscript. A.N.P. performed experiments and made simulations. F.S. helped with the experiments and contributed to the design of the investigation. J.-B.H. supervised the overall research effort. All authors reviewed the manuscript and suggested improvements.

**Graphical Abstract**

## Methods

**Materials.** The sample solution contained 2 mM 1,4-Bis(diphenylphosphino) butane (1,5-cyclooctadiene) rhodium(I) tetrafluoroborate (Strem Chemicals, CAS: 79255-71-3) and 60 mM ethyl phenylpropiolate (EP, Scheme 1A, Sigma-Aldrich, CAS: 2216-94-6) dissolved in acetone-$d_6$ 99.8% (Deutero GmbH, CAS: 666-52-4). Upon hydrogenation, ethyl cinnamate (EC) is formed (Scheme 1A).

**Experimental setup.** Experiments were carried out on a 600 MHz spectrometer (Bruker Avance II) with a cryogenically cooled probe (TCI) with Q=$v/\Delta v \cong$ 600.2 MHz/1.2 MHz $\cong$ 500 (see SI, Fig. S3) and 5 mm screw-cap NMR tubes (Wilmad). Tubes were filled with 500 μl of the sample solution. $p$H$_2$ was prepared using a home-build liquid nitrogen generator that provides a 50 % para enrichment. The gas was delivered into the spectrometer by a 1/16" polytetrafluoroethylene (PTFE) capillary. A hollow optical fiber was glued with epoxy resin to the end of the capillary and inserted into the NMR tube and solution (Molex, part. num. 106815-0026, internal diameter 250 μm and outer diameter 360 μm). A $p$H$_2$ pressure of approx. 1.2 bar (0.2 bar overpressure to atmosphere) was used to achieve a steady bubbling.

**Protocol 1 (Scheme 1b).** $p$H$_2$ was supplied (bubbled) to the sample solution for $\tau_b$ = 10 s to hydrogenate EP and generate PASADENA. After that, a rectangular, 4 μs 45° RF-pulse was applied. All experiments were carried out at 25 °C and ambient pressure. During the experiments, some convection and diffusion occurred. $p$H$_2$ bubbling was stopped only after RF-excitation and approx. 12.5 minutes of signal acquisition.

Continuous PHIP-RASER emission (Fig. 1B) was composed of 13 FID's of 1153844 points which corresponded to an acquisition time of 60 s per single FID. These FIDs were acquired sequentially with 1-3 s delay between experiments, this time was required for the spectrometer to load and start new acquisition. PHIP-RASER spectrum (Fig. 1D,E) is average of 11 spectra acquired that corresponds to FIDs №2-12, i.e. the first and the last FIDs were not used. The spectrum on Fig. 1E (blue) was smoothed using Savitzky-Golay filter with the first order polynomial for window of 100 points (~0.45 Hz or °0.75 ppb).

**Simulation.** Simulation parameters for Fig 1C and E are: initial density matrix $\hat{\rho}(-0) = \hat{1}/4 - \hat{I}_{1Z}\hat{I}_{2Z}$, equilibrium state $\hat{\rho}_{eq} = \hat{1}/4$, J = 10 Hz, chemical shift difference 1.014 ppm, B$_0$=14.1 T, $\alpha_{RD}$ = 1000 s$^{-1}$, relaxation time 5 s and rate of polarization influx $W_{in} = 0.5$ s$^{-1}$ during the first 400 seconds, and $0.5e^{-t0.2s^{-1}}$ s$^{-1}$ after that. Simulation model is described in SI and source code is available online.[23]

**Data availability.** Raw NMR data is provided in supplementary materials. Other data is available from the corresponding authors upon reasonable request.

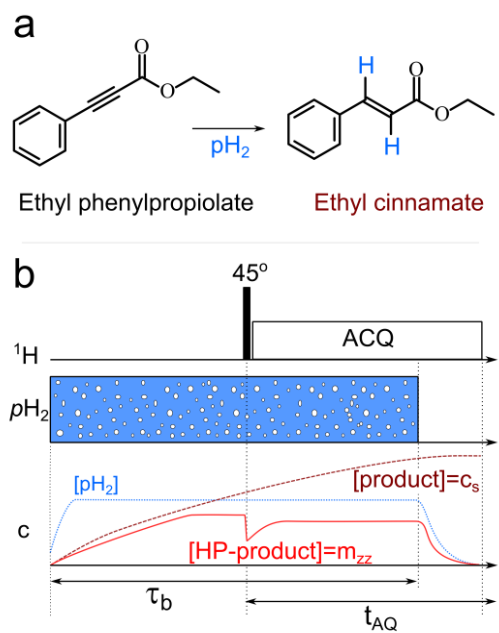

**Scheme 1.** **(A)** Schematic view of the reaction where ethyl phenylpropiolate (EP) is hydrogenated to ethyl cinnamate (EC). **(B)** Scheme of the experimental workflow (**protocol 1**): i) $p$H$_2$ supply for $\tau_b$. ii) excitation by hard 45° RF pulse and iii) signal acquisition (ACQ). Below, the concentrations of $p$H$_2$ (dotted blue line), total EC ([product]=c$_s$, dashed wine line) and hyperpolarized EC ([HP-product]=m$_{zz}$, solid red line) are plotted qualitatively. Note that $p$H$_2$ is immune to excitations with RF pulses and steady state conditions for m$_{zz}$ can be reached and different before and after RF excitation.

# Supporting Information

# Continuous Radio Amplification by Stimulated Emission using Parahydrogen Induced Polarization (PHIP-RASER) at 14 Tesla

Andrey N. Pravdivtsev,*[a] Frank D. Sönnichsen[b] and Jan-Bernd Hövener*[a]

[a] Dr. A. P. Pravdivtsev (0000-0002-8763-617X), Prof. Dr. rer. nat. Prof. J.-B. Hövener (0000-0001-7255-7252)
Section Biomedical Imaging, Molecular Imaging North Competence Center (MOIN CC), Department of Radiology and Neuroradiology,
University Medical Center Kiel, Kiel University (http://www.moincc.de/)
Am Botanischen Garten 14, 24118, Kiel, Germany
E-mails: andrey.pravdivtsev@rad.uni-kiel.de; jan.hoevener@rad.uni-kiel.de
[b] Prof. Dr. rer. nat. F. Sönnichsen (0000-0002-4539-3755)
Otto Diels Institute for Organic Chemistry, Kiel University, Otto Hahn Platz 5, 24098, Kiel, Germany
*correspondence to: andrey.pravdivtsev@rad.uni-kiel.de; jan.hoevener@rad.uni-kiel.de

**Table of content**

1. Evolution of PHIP-RASER emission (supplementary to Fig 1B)

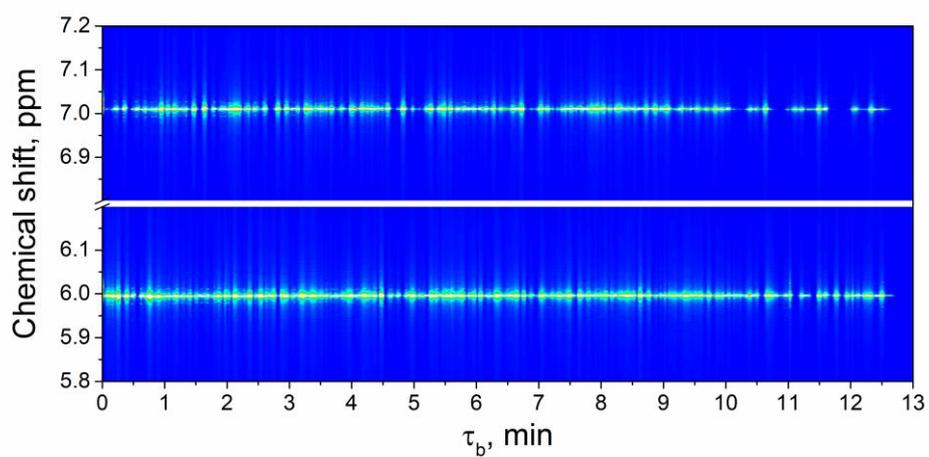

**Figure S1.** Observed evolution of two PHIP-RASER active lines (supplementary to Fig 1B): whole FID (13 minutes) was divided on blocks with duration of AQ=2 s, then Fourier transformation (FT) to each block was applied. Note that field inhomogeneity and convection due to $p$H$_2$ bubbling did not allow continuous stable RF-emission.

**2. NMR spectra acquired during and without *p*H₂ bubbling (supplementary to Fig 1)**

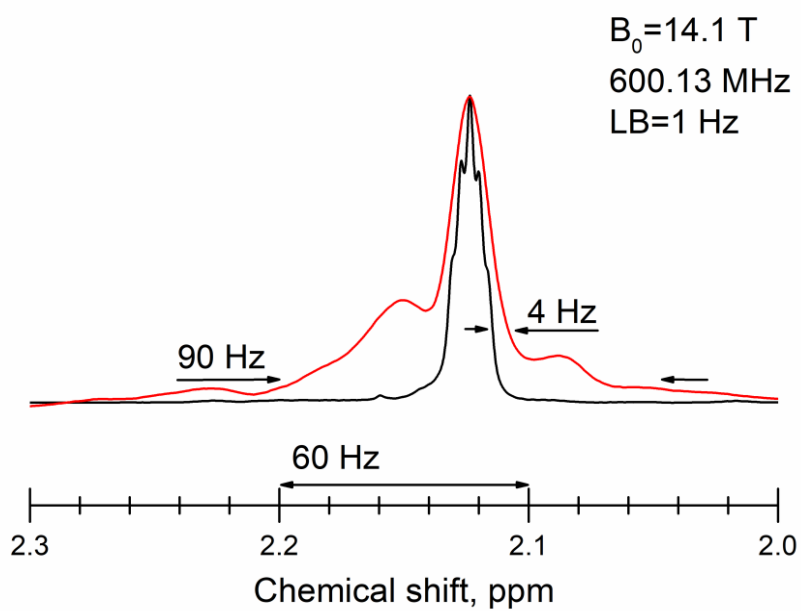

**Figure S2.** Homogeneous NMR spectrum (black) and spectrum acquired when *p*H₂ bubbling was on (red). Applied line broadening was 1 Hz. FID of the "red" spectrum is shown on Fig. 1B.

### 3. WOBB image of used NMR probe

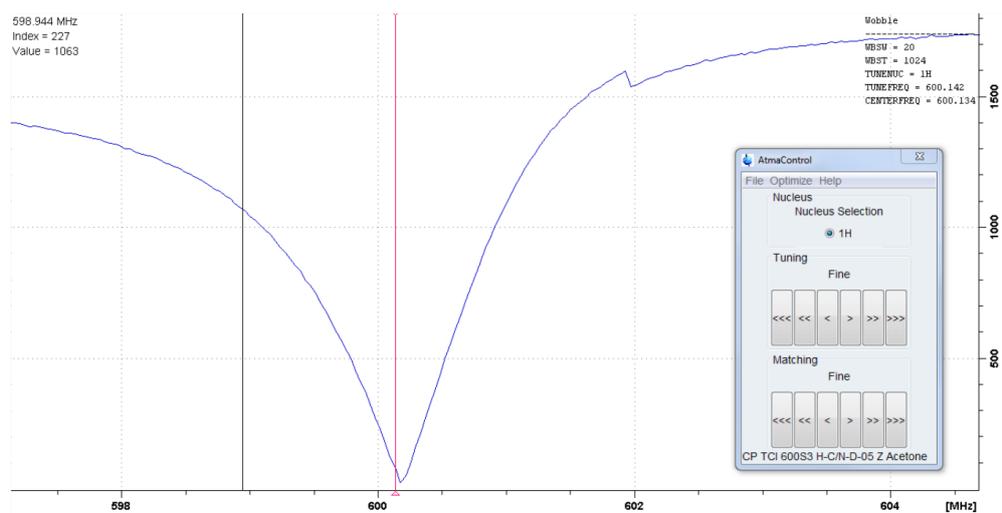

**Figure S3.** Image of a WOBB window in TopSpin for the used cryogenically cooled probe (TCI). It follows that Q=$v_0/\Delta v_{1/2}$ ≅600 MHz / 1.2 MHz ≅ 500.

## 4. Calculation of radiation damping time and other related parameters

Thermal polarization level and magnetization of spin-1/2 are given by[1]:

$$P = \tanh\left(\frac{\hbar \gamma B_0}{2k_B T}\right) \cong \frac{\hbar \gamma}{2k_B T} B_0 \ll 1 \qquad \text{(eq S1)}$$

and

$$M_0 = \frac{1}{2} \hbar \gamma n_s |P| \cong n_s \frac{\hbar^2 \gamma^2}{4 k_B T} B_0 \qquad \text{(eq S2)}$$

Then radiation damping rate, $(\tau_{RD})^{-1} = \alpha_{RD}|P|$ and parameter $\alpha_{RD}$ are given by the following equations[1]:

$$(\tau_{RD})^{-1} = \frac{1}{2} \mu_0 \eta Q |\gamma M_0| = \frac{\mu_0}{4} \hbar \gamma^2 \eta Q c_s |P| = \frac{\mu_0}{8 k_B T} \eta Q n_s \hbar^2 \gamma^3 B_0 \qquad \text{(eq S3)}$$

$$\alpha_{RD} = (\tau_{RD})^{-1}/P = \frac{\mu_0}{4} \hbar \gamma^2 \eta Q c_s \qquad \text{(eq S4)}$$

Filling factor, $\eta$ is usually in the range of 0.1-1 and is defined as follows[1]:

$$\eta \cong \frac{\text{volume of sample}}{\text{volume of coil}} \qquad \text{(eq S5)}$$

$$\frac{\mu_0}{4} \hbar \gamma^2 \eta Q c_s^0 \quad |P| * f$$

$$|P| * f = m$$

Below we will calculate the value of $P$, $M_0$ and $(\tau_{RD})^{-1}$ and $\alpha_{RD}$ for $H_2O$ at 300 K temperature and 10 Tesla magnetic field. High product $\eta Q = 100$ is assumed.

All necessary constants and values:

$\gamma_{1H} \cong 267.5 \cdot 10^6 \frac{rad}{s \cdot T}$ – gyromagnetic ratio of $^1H$ nuclei spin,

$\hbar \cong 1.054 \cdot 10^{-34} \frac{m^2 \cdot kg}{rad \cdot s}$ – Plank constant,

$c_s(H_2O) = 2c(H_2O) \cong 6.7 \cdot 10^{28} \frac{1}{m^3} = 6.7 \cdot 10^{22} \frac{1}{cm^3}$ – concentration of $^1H$ in pure $^1H_2O$,

$\mu_0 = 4\pi \cdot 10^{-7} \frac{T \cdot m}{A} \cong 1.257 \cdot 10^{-6} \frac{T \cdot m}{A}$ – vacuum permeability,

$k_B \cong 1.38 \cdot 10^{-23} \frac{m^2 \cdot kg}{s^2 \cdot K}$ – Boltzmann constant,

$[T] = \frac{kg}{A \cdot s^2}$ – Tesla units in the international system of units (known as SI).

Note that here we are keeping radian units for convenience.

Thermal polarization of $^1H$ at 300 K and 10 T:

$$P \cong \frac{\hbar \gamma B_0}{2 k_B T} \cong \frac{1.054 \cdot 10^{-34} \cdot 267.5 \cdot 10^6 \cdot 10 \frac{m^2 \cdot kg}{rad \cdot s} \cdot \frac{rad}{s \cdot T} \cdot T}{2 \cdot 1.38 \cdot 10^{-23} \cdot 300 \frac{m^2 \cdot kg}{s^2 \cdot K} K} = \frac{1.054 \cdot 10^{-27} \cdot 267.5}{2.76 \cdot 10^{-23} \cdot 300} \cong \cong 3.4 \cdot 10^{-5}.$$

Thermal magnetization of $^1H$ of $^1H_2O$ at 300 K and 10 T:

$M_0 = \frac{1}{2}\hbar\gamma n_s P \cong \frac{1}{2} 1.054 \cdot 10^{-34} \cdot 267.5 \cdot 10^6 \cdot 6.7 \cdot 10^{28} \cdot 3.4 \cdot 10^{-5} \frac{m^2 \cdot kg}{rad \cdot s} \cdot \frac{rad}{s \cdot T} \cdot \frac{1}{m^3} \cong$ $\cong$
$0.0322 \frac{kg}{s^2 \cdot m \cdot T}.$

Radiation damping rate of $^1$H of $^1$H$_2$O at 300 K and 10 T:

$(\tau_{RD})^{-1} = \frac{1}{2}\mu_0 \eta Q |\gamma M_0| \cong \frac{1}{2} 1.257 \cdot 10^{-6} \cdot 100 \cdot 267.5 \cdot 10^6 \cdot 0.0322 \frac{T \cdot m}{A} \cdot \frac{rad}{s \cdot T} \cdot \frac{kg}{s^2 \cdot m \cdot T} =$ $=$
$\frac{1}{2} 1.257 \cdot 2.675 \cdot 0.0322 \cdot 10^4 \frac{kg \cdot rad}{A \cdot s^3 \cdot T} \cong 540 \frac{rad}{s},$

and hence $\tau_{RD} \cong 1.8\ ms$,

Parameter $\alpha_{RD}$ for $^1$H of $^1$H$_2$O at 300 K and 10 T:

$\alpha_{RD} = \frac{\mu_0}{4}\hbar\gamma^2 \eta Q n_s = \frac{(\tau_{RD})^{-1}}{P} = \frac{540}{3.4 \cdot 10^{-5}} \frac{rad}{s} \cong 1.6 \cdot 10^7 \frac{rad}{s}$

Hence all together: $P \cong 3.4 \cdot 10^{-5}$, $M_0 \cong 0.0322 \frac{kg}{s^2 \cdot m \cdot T}$, $(\tau_{RD})^{-1} \cong 540 \frac{rad}{s}$, $\tau_{RD} \cong 1.8\ ms$ and $\alpha_{RD} \cong 1.6 \cdot 10^7 \frac{rad}{s}$ for $^1$H of $^1$H$_2$O at 300 K, 10 T and $\eta Q = 100$.

5. **Radiation damping model**

To describe the RD of a single nuclei, it is convenient to use modified Bloch-Maxwell equation proposed by Bloembergen and Pound that in the rotating frame of reference can be written as[2]

$$\frac{d}{dt}\begin{pmatrix}M_X\\M_Y\\M_Z\end{pmatrix}=\begin{pmatrix}-R_2 & \omega_Z & -\omega_Y\\-\omega_Z & -R_2 & \omega_X\\\omega_Y & -\omega_X & -R_1\end{pmatrix}\begin{pmatrix}M_X\\M_Y\\M_Z\end{pmatrix}+\frac{1}{M_0\tau_d}\begin{pmatrix}-M_XM_Z\\-M_YM_Z\\M_X^2+M_Y^2\end{pmatrix}+\begin{pmatrix}0\\0\\R_1M_0\end{pmatrix} \quad \text{(eq S6)}$$

$M_X, M_Y, M_Z$ are the components of magnetization vector $\mathbf{M}$ with Z being along magnetic field $B_0$. $M_0 = \frac{1}{2}\hbar\gamma c_s|P|$ is a value of equilibrium magnetization $\mathbf{M_0}=(0,0,M_0)$. $\hbar$ is Planck constant. $\gamma$ is gyromagnetic ratio. $c_s$ is the nuclei spin concentration and $P$ is the value of longitudinal polarization. $(\tau_d)^{-1}=\frac{\mu_0}{2}\eta Q|\gamma M_0|$ is the classical radiation damping rate,[1] where $\mu_0$ is a vacuum permeability, Q is the quality factor of the coil with the filling factor $\eta$. Using vector notations: $\mathbf{M}$, $\mathbf{M_0}$, $\mathbf{R}$ and $\boldsymbol{\omega}$ eq S6 can be simplified as

$$\frac{d\mathbf{M}}{dt}=[\mathbf{M}\times\boldsymbol{\omega}]-\mathbf{R}(\mathbf{M}-\mathbf{M_0})+\frac{1}{M_0\tau_d}[\mathbf{M}\times(-M_Y,M_X,0)] \quad \text{(eq S7)}$$

While Bloch equations are well suited to describe single spin systems, more complex effects are not included. To describe coupled multi spin systems, it is convenient to use Liouville von Neumann equation (LvN). However, a LvN equation that takes RD into account was not found in the literature. Using eq S7 and knowing the effect of different interactions on a spin system[3], we integrated RD into the following phenomenological LvN equation:

$$\frac{d}{dt}\widehat{\rho}(t)=-i\left(\widehat{\widehat{H}}+\widehat{\widehat{V}}^{rd}(t,\widehat{\rho})\right)\widehat{\rho}(t)-\widehat{\widehat{R}}\left(\widehat{\rho}(t)-\widehat{\rho}_{eq}\right)+\widehat{S} \quad \text{(eq S8)}$$

Where $\widehat{\rho}$ is a density matrix, $\widehat{\widehat{R}}$ is a relaxation superoperator, $\widehat{\rho}_{eq}$ is an equilibrium density matrix, $\widehat{\widehat{H}}=[\widehat{H},\cdot]$ is Hamiltonian superoperator of Hamiltonian $\widehat{H}$, and $\widehat{\widehat{V}}^{rd}=[\widehat{V}^{rd},\cdot]$ is a RD superoperator of RD operator $\widehat{V}^{rd}$. $\widehat{S}$ is a source of hyperpolarization (see below). For calculating relaxation superoperators $\widehat{\widehat{R}}$, we used a local fluctuating magnetic fields relaxation model, where relaxation of individual spin is defined by a single relaxation time at high magnetic field[4–6]. Then for multi spin system in the liquid state $\widehat{H}$ and $\widehat{V}^{rd}$ operators are

$$\widehat{H}=-\sum_k\boldsymbol{\omega}_k\cdot\widehat{\mathbf{I}}_k+2\pi\sum_{k<m}J_{km}\widehat{I}_{kZ}\widehat{I}_{mZ} \quad \text{(eq S9)}$$

$$\widehat{V}^{rd}(t,\widehat{\rho})=\alpha_{RD}\sum_k(m_{kY}(t)\,\widehat{I}_{kX}-m_{kX}(t)\widehat{I}_{kY}) \quad \text{(eq S10)}$$

Here $m_{kX,Y}=Tr\left(\widehat{\rho}\cdot\frac{1}{2}\widehat{I}^\dagger_{kX,Y}\right)/Tr\left(\frac{1}{2}\widehat{I}_{kX,Y}\cdot\frac{1}{2}\widehat{I}^\dagger_{kX,Y}\right)$ are amplitudes of polarization of of $\widehat{I}_{kX}$ and $\widehat{I}_{kY}$ components of spin $k$ of the density matrix $\widehat{\rho}$, the amplitudes in general are time dependent. In the main text we also showed evolution of polarization of double quantum longitudinal spin state calculated as $m_{ZZ}=Tr(\widehat{\rho}\cdot\widehat{I}_{1Z}\widehat{I}_{2Z})/Tr(\widehat{I}_{1Z}\widehat{I}_{2Z}\cdot\widehat{I}_{1Z}\widehat{I}_{2Z})$. $\alpha_{RD}=\frac{\mu_0}{4}\hbar\gamma^2\eta Qc_s=(\tau_{RD})^{-1}/|P|$ is a radiation damping rate without polarization factor. $\widehat{V}^{rd}$ can be understood in the same way as vector $(-M_Y,M_X,0)$ in eq S7 as induced by transverse magnetization electrical current. During hyperpolarization experiments, magnetization can change due to:

- variation of concentration of hyperpolarized agent, $c_S$, and
- variation of level of average polarization, $|P|$, due to e.g., relaxation and RF-excitation.

The rate of RD is linearly proportional to both parameters, which were not distinguished in the simulations; their product was represented by the corresponding amplitudes of density matrix $m_{kX,Y}$.

For the protons of water, with a concentration of ~110 M and a $\eta Q \cong 100$, $\alpha_{RD}=\frac{\mu_0}{4}\hbar\gamma^2\eta Qc_s \cong 10^7$ s$^{-1}$. In thermal equilibrium at 9.4 T, P($^1$H) is about $3.4\cdot 10^{-5}$ so that $(\tau_{RD})^{-1} \cong 540$ s$^{-1}$ or $\tau_{RD} \cong 1.8$ ms (see evaluation in section 4 above): this is an example of a very strong RD effect in thermal equilibrium. However, when the concentration is reduced by 5 orders of magnitude to 1 mM, then $\alpha_{RD} \cong 10^3$ s$^{-1}$ and at thermal equilibrium $\tau_{RD} \cong 180\ s \gg T_2^*$ hence RD can be completely neglected. On the other hand, when 1 mM sample is hyperpolarized to ~100 %, $\tau_{RD} \cong 6$ ms and radiation damping effect already play a crucial role.

The source of hyperpolarization, $\widehat{S}$, was introduced into eq. S8 to mimic the experimental conditions. A similar approach was used before e.g., to model SABRE[7,8] and DNP[9]. During the experiment, the hydrogenation reaction carries on, corresponding to an influx of para spin order into spin system. For a two spin-1/2 system and enrichment of *p*H$_2$ to 100 %, the source operator can be written as

$$\widehat{S} = W_{in}(t)\left[-\left(\hat{\mathbf{I}}_1 \cdot \hat{\mathbf{I}}_2\right)\right] \qquad \text{(eq S11)}$$

Where $-\left(\hat{\mathbf{I}}_1 \cdot \hat{\mathbf{I}}_2\right)$ is the traceless part of the *p*H$_2$ density matrix, and $W_{in}$ is the rate of hyperpolarization income (generally time dependent). We used a constant value and an exponential decay function (Fig 1 and S8) and a superposition of delta functions (Fig S6) to construct $W_{in}(t)$. Simulations were performed using the MOIN spin library[8,10]; the source code is available online.

6. *Simulation:* **radiation damping in the case of thermal equilibrium**

To test introduced above RD model, we simulated a thermally polarized, weakly coupled two spin-1/2 system. The simulated RD effect was similar to that observed experimentally: transversal and longitudinal magnetization were depleted or restored faster, respectively (Fig. S4B). All spectral components were equally broadened (Fig. S4D).

These RD effects are well known,[11] but calculated for the coupled multi spin system using density matrix approach for the first time. In the same way, RD effects can be simulated for any NMR pulse sequences and coupled multi spin systems.

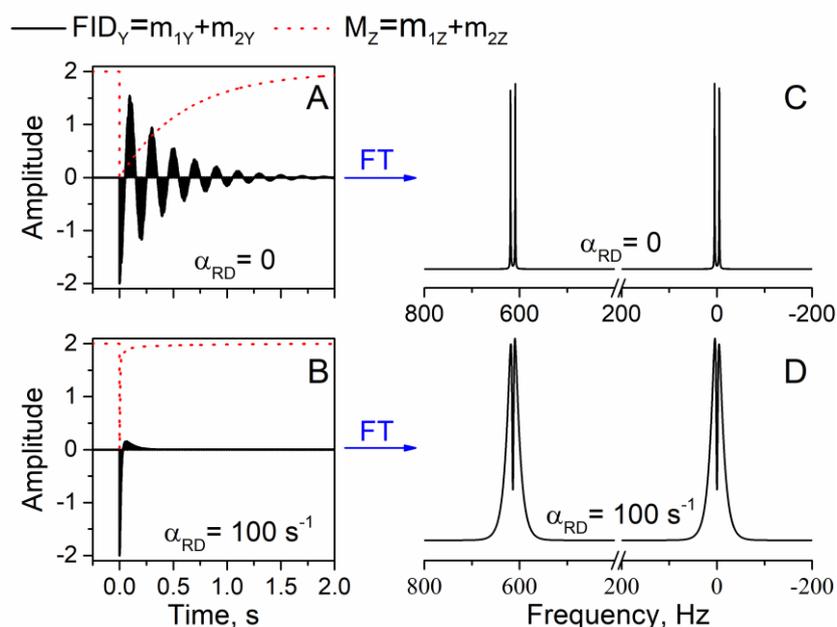

**Figure S4.** Simulated free induction decay (FID$_Y$=$m_{1Y} + m_{2Y}$), longitudinal magnetization ($M_Z = m_{1Z} + m_{2Z}$) (A, B) and corresponding NMR spectra (C, D) for an AX spin system with RD (B, D: $\alpha_{RD}$=100 s$^{-1}$) and without RD (A, C: $\alpha_{RD}$=0). A completely polarized system, $\hat{\rho}(-0) = \hat{\rho}_{eq} = \hat{1}/4 + (\hat{I}_{1Z} + \hat{I}_{2Z})/2$, was assumed. At time point 0 ideal $90°_{-X}$ rotation pulse was applied. Simulation parameters: B$_0$=14.1 T, J = 10 Hz, $W_{in}$=0, chemical shift difference 1.023 ppm and relaxation time 0.6 s.

7. *Simulation:* **effect of radiation damping on PASADENA**

These simulations (Fig. S5) show the difference of PASADENA dynamic between cases when radiation damping can be neglected and when it is much faster than $T_2$ relaxation rate.

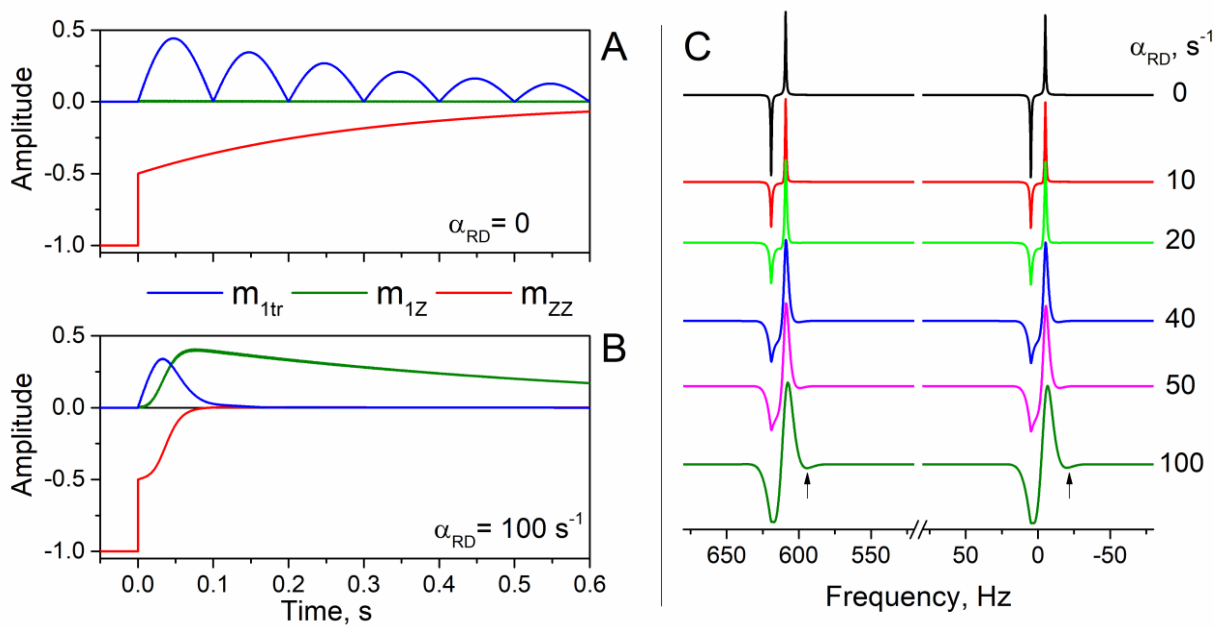

**Figure S5.** Simulations of PASADENA experiment with varying rate of radiation damping: evolution of transversal ($m_{1tr} = \sqrt{m_{1X}^2 + m_{1Y}^2}$) and longitudinal ($m_{1Z}$) magnetizations and $m_{ZZ}$ (amplitude of $\hat{I}_{1Z}\hat{I}_{2Z}$ spin order) for $\alpha_{RD} = 0$ (no RD, A) and $\alpha_{RD} = 100$ s$^{-1}$ (B), and corresponding NMR spectra for $\alpha_{RD} = (0 - 100)$ s$^{-1}$ (C). After 45° RF-excitation RD transforms remaining multiplet ZZ-polarization into net magnetization: $\hat{I}_{1Z}\hat{I}_{2Z} \xrightarrow{RD} \hat{I}_{1Z} + \hat{I}_{2Z}$ (B), and also causes asymmetric line broadening (C). Appearance of extra deeps indicated by arrows. Initial density matrix is $\hat{\rho}(-0) = \hat{1}/4 - \hat{I}_{1Z}\hat{I}_{2Z}$. In equilibrium system is not polarized: $\hat{\rho}_{eq} = \hat{1}/4$. Simulation parameters are J = 10 Hz, chemical shift difference 1.023 ppm, B$_0$=14.1 T, $W_{in}$=0 and relaxation time 0.6 s.

8. *Simulation:* **spontaneous emission in PASADENA experiment. Production of hyperpolarization is represented by superposition of delta functions**

This simulation shows that after the introduction of polarization the system needs some time before it starts to emit.

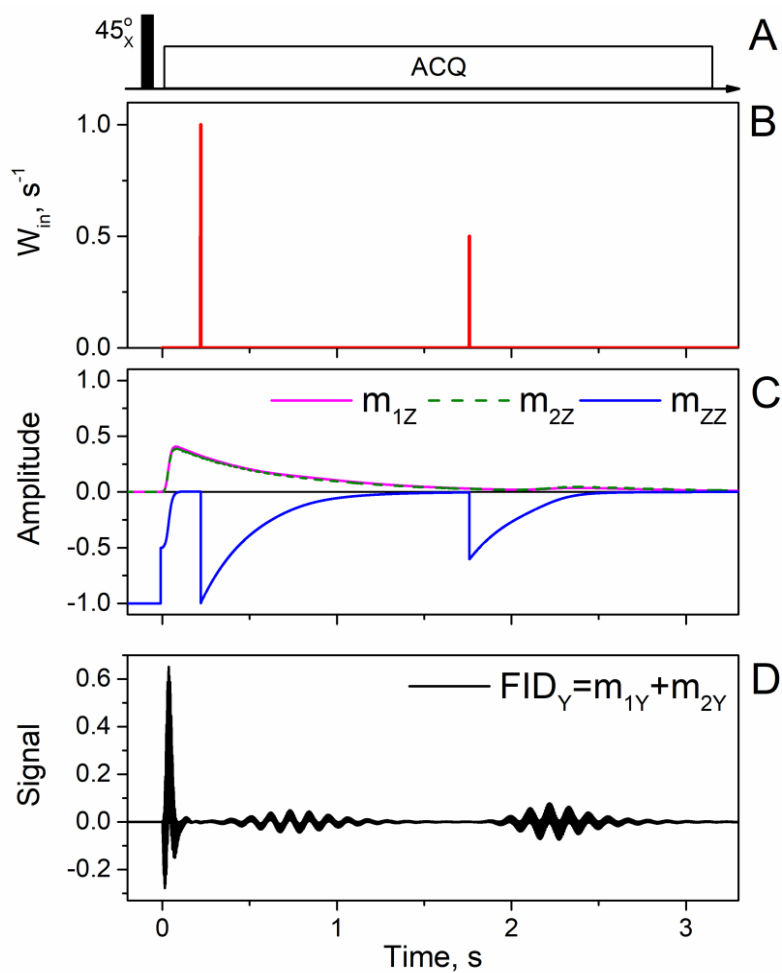

**Figure S6.** PASADENA simulations with spontaneous emission of "echo" like signals after sudden injection of polarization: sequence diagram (A), the hyperpolarization production rate, $W_{in}$ represented with two delta functions (B), evolution of amplitudes of net $m_{1Z}$ and $m_{2Z}$ and multiplet $m_{ZZ}$ magnetization as a function of time (C) and FID in corresponding PASADENA experiment (D). Radiation damping together with multiplet polarization results in "spontaneous" emission of echo like signals. Simulation parameters are: initial density matrix $\hat{\rho}(-0) = \hat{1}/4 - \hat{I}_{1Z}\hat{I}_{2Z}$, equilibrium state $\hat{\rho}_{eq} = \hat{1}/4$, J = 10 Hz, chemical shift difference 1.023 ppm, B$_0$=14.1 T; $\alpha_{RD}$ = 100 s$^{-1}$, relaxation times 0.6 s and operator of polarization source $\hat{S} = W_{in}(t)[-\hat{I}_{1Z}\hat{I}_{2Z}]$. Depending on model parameters: relaxation rate, $\alpha_{RD}$, $W_{in}(t)$ different number and shape of emitted signals can be obtained. Here we demonstrate that two instant additions of PASADENA polarization in the presence of weak transverse magnetization results in two "echo" like emitted signals.

## 9. Experimental protocol 2: stop bubbling scheme

**Protocol 2 (Scheme S1).** $pH_2$ was supplied (bubbled) to the sample solution for $\tau_b$ = 5 s to hydrogenate EP and generate PASADENA. Next, the bubbling was stopped and the system was allowed to settle for a variable period of time, $\tau_w \geq 2$ s. After that, a rectangular 4 µs 45° RF-pulse was applied. Because $pH_2$ is immune to RF pulses, hydrogenation reaction of EP with remaining dissolved $pH_2$ continues after RF-excitation. This experimental procedure allowed us to generate hyperpolarization after RF-pulse keeping high magnetic field homogeneity. All experiments were carried out at 25 °C and ambient pressure. During the experiments, some convection and diffusion occurred; the effect was neglected.

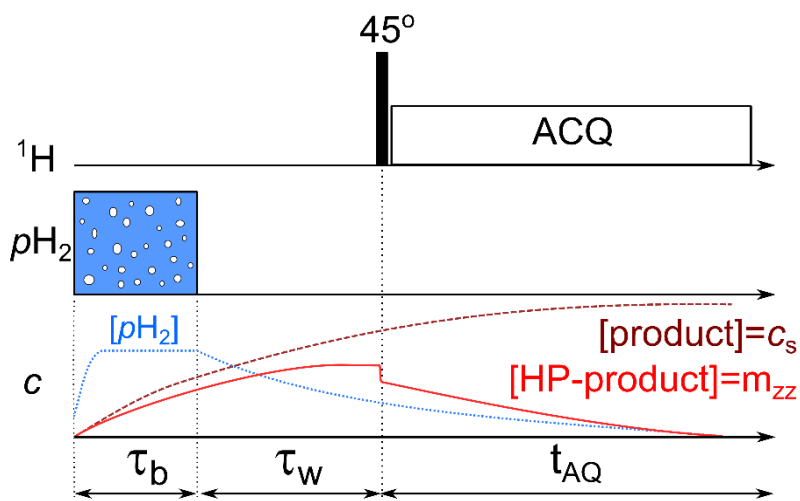

**Scheme S1.** Scheme of the experimental workflow (**protocol 2**): i) $pH_2$ supply for $\tau_b$ = 5 s with 0.2 bar overpressure. ii) waiting period 2 s < $\tau_w$ < 20 s; iii) excitation by hard 45° RF pulse and iv) signal acquisition (ACQ). Below, the concentrations of $pH_2$ (dotted blue line), total EC ([product]=$c_s$, dashed wine line) and hyperpolarized EC ([HP-product]=$m_{zz}$, solid red line) are plotted qualitatively. Note that hydrogenation with $pH_2$ continues after the RF excitation ($pH_2$ is immune to excitations with RF pulses).

## 10. Spontaneous PHIP-RASER bursts (protocol 2)

Interestingly, when RD is active, $\hat{I}_{1Z}\hat{I}_{2Z}$ spin order that remains after excitation, is quickly converted into longitudinal magnetization: $\hat{I}_{1Z}\hat{I}_{2Z} \xrightarrow{RD} \hat{I}_{1Z} + \hat{I}_{2Z}$ (Fig S7D [0,0.1 s]). Transverse magnetization and resulting longitudinal magnetization relaxes with $\sim\tau_{RD}$ and $\sim T_1$ times respectively. The fast decay of transverse magnetization reveals itself in a broadening of the spectral lines, which is in fact asymmetric: emission lines are broader and less intense than absorption lines (Fig S7C,E). When this asymmetry is observed in the experiment, it is a first indication of the presence of RD in PASADENA experiments. If undesired, special care should be taken, e.g. detune probe (reduce Q), reduce field homogeneity or use pulse sequences with dephasing gradients.[12,13]

Another, even more important aspect occurs later. If $p$H$_2$ is still present after 45° RF-excitation, then hydrogenation reaction and hence production of PASADENA polarization continuous.

We found strong "spontaneous" bursts of RASER with an echo-like appearance up to 30 s after finishing $p$H$_2$ bubbling and consequent excitation with 45° RF-pulse (Fig S7B). The echo like signals manifest itself as a very narrow "artifacts" on an emission spectral lines (compare Fig. S7C-8 and 9).

The experiment was repeated >20 times without changing the sample, and the observations were reproduced. However, during the experiment the concentration of substrate and the activity of catalyst are changing. This results in differences in the number of appearing echo like signals and their positions.

The observations were qualitatively reproduced by simulating a coupled spin system with RD and source of hyperpolarization (Fig S7D,E). Transverse magnetization via RD excites multiplet spin order of PASADENA ($\hat{I}_{1Z}\hat{I}_{2Z}$) and converts it into longitudinal and transverse magnetization. The latter manifests itself as a "spontaneous" emission of echo like signal. The amplitude of emitted signals decay from burst to burst because $p$H$_2$ is consumed and hence polarization production rate decays.

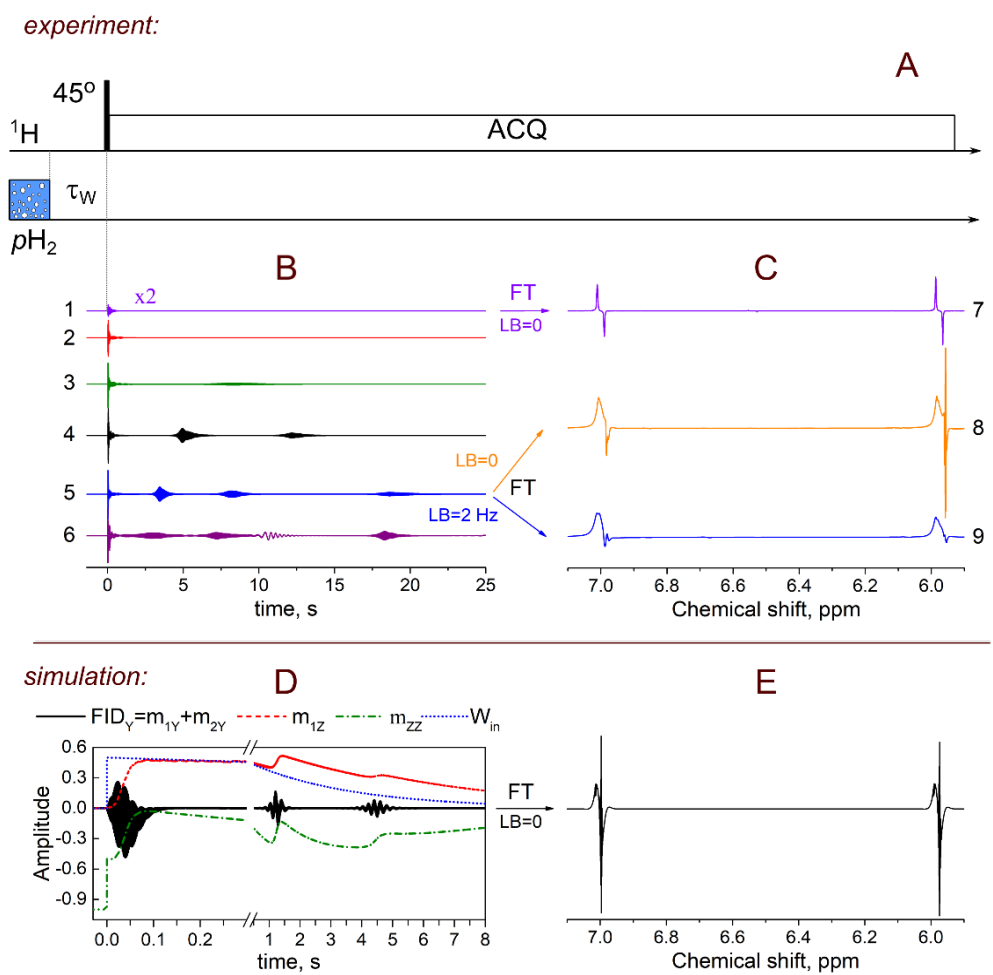

**Figure S7.** PASADENA experiment with spontaneous emission of "echo" like signals: sequence diagram (A), experimental observations (B), simulated evolution of PASADENA ($m_k$ is polarization level of spin order $k=\hat{I}_{1Y}$, $\hat{I}_{2Y}$, $\hat{I}_{1Z}$ and $\hat{I}_{1Z}\hat{I}_{2Z}$) (D) and corresponding NMR spectra (C and E). Six ¹H NMR free induction decays (FIDs) were acquired after 45° excitation of the sample that was presaturated with $pH_2$ each time before RF-excitation. Concentration of substrate and catalyst activity decreases from (6) to (2), $\tau_w = 1$ s. FID (1) was multiplied by a factor of 2 for a better visual perception, $\tau_w = 25$ s. Spectra (8-9) are Fourier transformation (FT) of FID (5) with line broadening (LB) 0 and 2 Hz respectively. Note the "spontaneous" emission of PASADENA and RD is more profound on emission spectral components; absorption components are predictably broadened. Simulation parameters are: initial density matrix $\hat{\rho}(-0) = \hat{1}/4 - \hat{I}_{1Z}\hat{I}_{2Z}$, equilibrium state $\hat{\rho}_{eq} = \hat{1}/4$, J = 10 Hz, chemical shift difference 1.023 ppm, $B_0$=14.1 T, $\alpha_{RD}$ = 100 s⁻¹, relaxation time 5 s and rate of polarization influx $W_{in} = 0.5e^{0.3s^{-1}\cdot t}$. Depending on model parameters: relaxation rate, $\alpha_{RD}$, $W_{in}(t)$ different number and shape of emitted signals can be obtained.